\documentclass[prl,twocolumn,showpacs]{revtex4}
\usepackage{graphicx}
\usepackage{amsmath}

\begin{document}

\title{Dynamical electron transport through a nanoelectromechanical wire in a magnetic field}
\author{Hangmo Yi$^1$ and Kang-Hun Ahn$^2$}
\affiliation{$^1$Department of Physics,
Soongsil University, Seoul 156-743, Korea \\
$^2$Department of Physics,
Chungnam National University, Daejeon 305-764, Korea}
\date{\today}

\begin{abstract}
We investigate dynamical transport properties of interacting
electrons moving in a vibrating nanoelectromechanical
wire in a magnetic field. We have
built an {\em exactly solvable} model in which electric 
current and mechanical oscillation are treated fully quantum
mechanically on an equal footing. Quantum mechanically fluctuating
Aharonov-Bohm phases obtained by the electrons cause nontrivial
contribution to mechanical vibration and electrical conduction
of the wire.
We demonstrate our theory by calculating the admittance of 
the wire which are influenced by the multiple interplay between
the mechanical and the electrical energy scales, magnetic field
strength, and the electron-electron interaction.
\end{abstract}

\maketitle

Recent progresses in experimental techniques of nano structures have
made it possible for researchers to investigate electromechanical
systems with extremely small sizes.\cite{bishop2001a}
The size limit is now being pushed down to the vicinity of
mechanically quantum regime and there are much effort to understand
the effect of quantum fluctuations on electrical and mechanical
properties --- and their coupling --- in the nanoelectromechanical
systems.\cite{blencowe2004a,ahn2004a,knobel2003a,schwab2005a,lahaye2004a,leroy2004a,sapmaz2006a}
Recently Shekhter and his coworkers investigated an Aharonov-Bohm effect in quamtum mechanically vibrating nanoelectromechanical wires.\cite{shekhter2006a} 
In the limit where non-interacting electrons tunnel between weakly connected leads through virtual processes, they used a perturbative analysis to find that the quantum interference has a strong effect on the electromechanical response of the wire. 

In this Letter, we propose an {\it exactly solvable} theoretical model of strongly interacting electrons in a nano wire which quantum-mechanically oscillates in a magnetic field. 
We study the dynamical electron transport beyond perturbative regime and investigate the fundamental limit of both quantum mechanics and strong correlations. 
Whereas Ref.~\onlinecite{shekhter2006a} focuses on energy scales much less than the level spacing of the vibrational normal modes, our model is valid for a much broader range of energy scales, including energies much greater than the vibrational energy level spacing. 
Therefore, it captures various important and interesting physical consequencies, such as admittance resonances occuring near mechanical normal mode frequencies.
We also consider the electron-electron interaction in the theory, which is known to play an important role in low dimensional systems such as carbon nanotubes.\cite{yao1999a} 
Our theory is based on a Luttinger liquid model which was previously developed by the present authors.\cite{ahn2004a}
One of the important characteristics of it is that it treats both electrical and mechanical degrees of freedom quantum mechanically, completely on an equal footing. 
On the atomic level, the interplay between electrical charge and mechanical oscillation often causes rich and interesting new physics to emerge --- the most famous examples being the BCS theory of superconductivity\cite{bardeen1957a} and the SSH theory of conducting polymers.\cite{su1979a}
This Letter deals with a mesoscopic version of such a case, in which one can even control the electromechanical coupling by tuning external parameters. 

Figure \ref{fig:setup} schematically shows the experimental setup of our model. 
A one-dimensional wire --- a single-wall carbon nanotube, for example --- of length $L$ is suspended across a valley between two metallic gates.
There are two important degrees of freedom for this model: the mechanical oscillation of the wire and the electric current through it.
Mechanically, the wire may form standing waves of which sound velocity is determined by its one-dimensional mass density and tension. 
When the amplitude of the oscillations is small, the mechanical motion of the wire may be characterized by its displacement $u(x)$ from the equilibrium position, where $x$ is the one-dimensional position along the wire.
By connecting the gates to an external ac voltage source, one may now drive an ac current through the wire.
If there is a magnetic field $\mathbf{B}$ perpendicular to the wire, the wire is expected to feel the Lorentz force and move.
This may in turn induce electromotive force in the wire as it cuts through the magnetic flux.
This is the well-known effect of back-reaction.\cite{ahn2004a}
We will discuss this effect in the regime where both the electron current and the mechanical oscillations must be treated quantum mechanically. 

The electronic excitations of our one-dimensional electromechanical system is best described by the Tomonaga-Luttinger liquid theory.\cite{tomonaga1950a,luttinger1963a,haldane1981b}
In this theory, the electron-electron interaction cannot be treated perturbatively, but the Hamiltonian may be rendered quadratic by a bosonization technique. 
This is the key to the exact solvability of our model. 
For simplicity, we will consider a spinless case here, but it may be easily generalized to include spin.
The electronic part of the Euclidean action is given by\cite{kane1992a,yi2002a}
\begin{equation}
  \!S_\theta  = \frac{1}{8\pi v_F} \int_0^\beta \!d\tau \int_{-\infty}^\infty dx \left\{ \left[\frac{\partial\theta}{\partial\tau}\right]^2 + \left[\frac{v_F}{K(x)}\frac{\partial\theta}{\partial x}\right]^2 \right\}
\end{equation}
where $v_F$ is the Fermi velocity, $\tau=it$ is the imaginary time, and $\beta\equiv 1/k_BT$ is the inverse temperature.
Here, $\theta(x)$ is a bosonic field related to the one-dimensional electronic density fluctuation $\delta n$ via $\delta n=(\partial\theta/\partial x)/2\pi$.
We have replaced a system of finite-sized one-dimensional Tomonaga-Luttinger liquid wire connected to two-dimensional leads by an effective one-dimensional liquid with position dependent interaction parameter and velocity:\cite{maslov1995a}
\begin{equation}
K(x) = \left\{
  \begin{array}{ll}
    K, & \text{if } 0<x<L, \\
    1, & \text{if } x<0 \text{ or } x>L. \\
  \end{array} \right.
\end{equation}
For a short range interaction of the form $V(x-x')=V\delta(x-x')$, we have $K=1/\sqrt{1+V/\pi v_F}$.
Inside the wire, the velocity of the acoustic plasmon is also renormalized to $v=v_F/K$.
In a more realistic interface, $K(x)$ would change more smoothly, but it will not make qualitative changes to the results below.

The mechanical part of the action is
\begin{equation}
  S_u = \frac{\rho}{2} \int_0^\beta d\tau \int_0^L dx \left[ \left(\frac{\partial u}{\partial\tau}\right)^2 + \left(v_s\frac{\partial u}{\partial x}\right)^2 \right]
\end{equation}
where $u(x)$ is the transverse displacement of the wire from the equilibrium position, $\rho$ the one-dimensional mass density of the wire, and $v_s$ the sound velocity of the mechanical transverse waves.

Finally, the two fields $\theta$ and $u$ are coupled via a magnetic field $\mathbf{B}=B\hat{\mathbf{z}}$.
We will choose a guage in such a way that the vector potential is given by $\mathbf{A} = -By\hat{\mathbf{x}}$.
At a given position along the wire, one may simply replace $y$ with the transverse position of the wire $u(x)$. The coupling part of the action is then given by
\begin{align}
  S_{\theta\text{-}u} & = \frac{1}{c} \int_0^\beta d\tau \int_0^L dx JA_x \nonumber \\
  & = \frac{1}{c} \int_0^\beta d\tau \int_0^L dx \left( \frac{-ie}{2\pi} \frac{\partial \theta}{\partial \tau} \right) (-Bu),
\end{align}
where $-e$ is the electron charge and $J=(-e/2\pi)(\partial \theta/\partial t)$ is the elctric current.
Since the total Action
\begin{equation}
  S = S_\theta + S_u + S_{\theta\text{-}u}
  \label{eq:action}
\end{equation}
is completely quadratic, this model is exactly solvable.

Before going into any detailed calculations, we would like to first discuss the qualitative characteristics of this model.
In the absence of magnetic field, $S_{\theta\text{-}u}=0$ and the electrical and mechanical degrees of freedom are completely decoupled. 
Then there are two independent branches of excitations.
The normal modes of the mechanical degree of freedom $u(x)$ are standing waves $u_m(x) = u_{m0}\sin(m\pi x/L)$, the energies of which are quantized as $m\Delta_u$ with an integer $m$. 
Here, $\Delta_u\equiv\pi\hbar v_s/L$ is the energy level spacing of the mechanical normal modes.
On the other hand, the electrical degree of freedom $\theta(x)$ is infinitely extended and its energy is not quantized.
However, due to the facts that the voltage drop occurs within the length of the wire and $K(x)$ changes at the boundaries, the energy scale $\Delta_\theta\equiv\pi\hbar v/L$ may still manifest itself in some physical quantities as will be shown below.
This in fact is the energy level spacing of the Tomonaga-Luttinger acoustic plasmon confined in a box of length $L$.

If the magnetic field is now turned on, the two degrees of freedom become coupled to each other through the following two processes: (i) a current induces a Lorentz force on the wire and (ii) an oscillation causes electromotive force through magnetic induction.
Therefore, stimulating one of the two degrees of freedom will induce excitations in another. 
Eventually, the effect will negatively feed back into the originally stimulated degree of freedom; this is the origin of the back-reaction.
What characterizes the coupling strength is the magnetic back-reaction energy scale $\omega_B\equiv B\sqrt{v_Fe^2/\pi\hbar\rho c^2}$.
This is the size of the back-reaction gap in one of the two uncoupled excitation branches.\cite{ahn2004a}
For a small magnetic field, the coupling is weak and the electronic excitations are only slightly perturbed.
The effect of the coupling will be most strong at each quantized energy level of the mechanical oscillations, which, in the weak field limit, occurs at every $\Delta_u$.
(In fact, even multiples of $\Delta_u$ will have no effect due to the reason explained below.)
As the magnetic field increases, the coupling grows stronger and the two branches of excitations intricately merge together to form new sets of excitations.
The dispersion relation for the new sets of excitations have been calculated for an infinitely long wire in Ref.\ \onlinecite{ahn2004a}.
In the strong coupling limit ($\omega_B\gg\omega\Delta_\theta/\Delta_u$), the dispersion relations are approximately given by $E(q) \approx v_svq^2/\omega_B$ and $\sqrt{v^2q^2+\omega_B^2}$.
As we will see below, these will characterize the energy profile of finite wires, too.

One of the important physical quantities that characterize the system is the current response to an external ac bias voltage.
Let us consider the average current inside the wire $I\equiv\int_0^L J(x)dx$.
The external bias voltage and the current will be assumed to take the form $V_\mathrm{ext}=V_0(\omega)e^{i\omega t}$ and $I=I_0(\omega)e^{i\omega t}$, respectively.
Then, we may calculate the ratio of the complex amplitudes of the current and the external bias voltage in the linear response regime, i.e., $Y(\omega)\equiv\lim_{V_0(\omega)\rightarrow 0} I_0(\omega)/V_0(\omega)$, which is the inverse of the impedance and is simply called the admittance.
For small biases, the linear response theory dictates
\begin{equation}
  Y(\omega) = \frac{1}{i\hbar\omega L^2} \int_0^L dx \int_0^L dx' \int_0^\infty dt e^{i\omega t} \Pi(x,x',t),
\end{equation}
where
\begin{equation}
\Pi(x,x',t) \equiv \left\langle J(x,t)J(x',0) \right\rangle = \left(\frac{e\omega}{2\pi}\right)^2 \left\langle \theta(x,t)\theta(x',0) \right\rangle
\end{equation}
is the current-current correlation function.
This may be computed using the Euclidean action in Eq.\ (\ref{eq:action}) and the usual analytic continuation technique.\cite{kane1992a}

The real part of the admittance is proportional to the power absorption spectrum.
Our calculations show that
\begin{equation}
  \mathrm{Re}\ Y(\omega) = \frac{e^2}{h} \frac{A(\omega)}{\omega^2+[\Gamma(\omega)]^2} \label{eq:ReY}
\end{equation}
where
\begin{align}
  \Gamma(\omega) & = \frac{\Delta_\theta}{2K\hbar}\left\{ [1-\kappa(\omega)]\lambda_+(\omega)\tan\frac{\pi\lambda_+(\omega)}{2} \right. \nonumber \\
    & \qquad\qquad + \left. [1+\kappa(\omega)] \lambda _-(\omega)\tan\frac{\pi\lambda_-(\omega)}{2} \right\} \\
  A(\omega) & = \frac{\omega^2}{\pi} \left[ \frac{1-\kappa(\omega)}{\lambda_+(\omega)}\tan\frac{\pi  \lambda_+(\omega)}{2} \right. \nonumber \\
  & \qquad\quad + \left. \frac{1+\kappa(\omega)}{\lambda_-(\omega)}\tan\frac{\pi\lambda_-(\omega)}{2} \right]^2
\end{align}
with
\begin{align}
  \kappa(\omega) & = \frac{1-\eta ^2}{\sqrt{\left(1-\eta^2\right)^2+4(\eta\omega_B/\omega)^2}}, \\
  \lambda_\pm(\omega) & = \frac{\hbar\omega}{\Delta_u}\left[\frac{1+\eta ^2}{2} \pm \sqrt{\left(\frac{1-\eta ^2}{2}\right)^2+\left(\frac{\eta\omega_B}{\omega}\right)^2}\right]^\frac{1}{2} \\
  \eta & \equiv \Delta_u/\Delta_\theta = v_s/v.
\end{align}
These quantities need to be evaluated numerically. 

Figure \ref{fig:Y(w)} shows the main result.
Assuming that the wire is a carbon nanotube, we have taken the value $K=0.22$ from Ref.\ \onlinecite{yao1999a}.
Since the mechanical sound velocity is usually small compared to the electron Fermi velocity, we have also assumed $\eta=\Delta_u/\Delta_\theta=v_s/v=0.1$.
For $\omega_B=0$, the mechanical degree of freedom is decoupled from the electronic one and does not contribute to the current measurement at all.
Therefore, $\Delta_\theta$ is the only relevant energy scale and it determines the frequency scale over which $\mathrm{Re}\,Y(\omega)$ decays at small $\omega$.
It also determines the period of weak modulation in the admittance, which is usually too small to see in the figures.
Note that the admittance approaches the usual Landauer dc conductance $e^2/h$, as $\omega\rightarrow 0$.\cite{maslov1995a}
This dc conductance is unaffected by magnetic field and stays unchanged for all values of $\omega_B$, simply because there is no back-reaction for a time-independent dc current.

For a weak magnetic field($\omega_B\ll\omega/\eta$), regularly placed sharp resonance peaks start to appear.
Their positions are determined by the condition $\Gamma(\omega)=0$.
Note that they occur whenever the frequency $\omega$ is close to an {\it odd} integer multiple of $\Delta_u/\hbar$.
From the positions of the resonances, we can deduce that the changes in the mechanical excitation energy levels are perturbatively small in the weak coupling limit.
The widths of the peaks are usually very small, with Q factors sometimes reaching as large as $\sim 10^7$.
They increase with the magnetic field as $B^2$, which may be easily understood as broadening effect due to the electromechanical coupling.
There is no resonance at even integer multiple of $\Delta_u/\hbar$ due to the following reason. 
The magnetic flux swept by the $m$th normal mode is $B\int_0^L u_m(x) dx$, but this always vanishes for an even integer $m$ because the areas above and below the wire cancel each other.

As the magnetic field grows, the coupling becomes stronger.
The positions and shapes of the resonance peaks change much and some peaks even get wiped out.
In the limit of strong magnetic field($\omega_B\gg\omega/\eta$), there are well distinguishable sharp peaks at $\omega\approx m^2\Delta_\theta\Delta_u/\hbar\omega_B$. 
This is in good agreement with the dispersion relation for the lower-energy gapless branch of the strongly coupled system with $q=m\pi/L$.\cite{ahn2004a}
If $\hbar\omega_B\gg\Delta_\theta,\Delta_u$, there are also broad peaks that develop near $\omega\sim\omega_B$, which is a result of crossover between the weak and strong coupling regimes.

Now let us discuss about the effect of the electron-electron interaction.
From Eq.\ (\ref{eq:ReY}), it is easy to see that $\mathrm{Re}\,Y(\omega)$ depends on $K$ only through an overall coefficient of $\Gamma(\omega)$.
In general, $\mathrm{Re}\,Y(\omega)$ decreases as the interaction becomes stronger, except right on resonance where $\mathrm{Re}\,Y(\omega)$ becomes independent of the interaction strength because $\Gamma(\omega)=0$.
Well away from resonances[$\Gamma(\omega)\gg\omega$], the value of $\mathrm{Re}\,Y(\omega)$ is approximately proportional to $K^2$.

All quantities discussed so far have no temperature dependence, because the action of our theory is completely quadratic.
It will not be true if, for example, the contacts between the wire and the leads are not perfectly transmitting.
In that case, the action will no longer be quadratic and there will be nonvanishing thermal effects in general. 
For partially transmitting contacts, we also have to consider the effect of charge quantization and Coulomb blockade, although carefully adjusting a backgate voltage may lift Coulomb blockade.\cite{kane1992b}
Although Ref.~\onlinecite{shekhter2006a} studies an opposite limit of weakly coupled leads, the results agree with ours in the sense that the electron transmission is depressed by magnetic fields, because it is a direct consequence of the back-reaction. 

In summary, we have shown that the quantum mechanically coupled electronic and mechanical degrees of freedom of a one-dimensional wire oscillating in a mangetic field may be studied in an exactly solvable model.
Calculating the admittance, we have found that the interplay of three important energy scales, the oscillation energy level spacing $\Delta_u$, the Tomonaga-Luttinger liquid energy level spacing $\Delta_\theta$, and the magnetic energy $\omega_B$ may result in rich and interesting consequencies such as sharp resonance peaks and crossover in terms of relative magnitude of the frequency to the magnetic induction coupling strength.
From the positions and shapes of the resonance peaks, we may extract such pieces of information as the magnetic field, electron-electron interaction, and the standing wave energy levels, which may be used in applications such as nano sensors.

This work was supported by the Korean Research Foundation Grant funded by
the Korean Government (MEST, Basic Research Promotion Fund,
KRF-2007-331-C00110) (H.Y.) and the Korea Science and Engineering
Foundation(KOSEF) grant funded by the Korean government (MEST, No.
R01-2007-000-10837-0) (K.H.A.).

\begin{center}
\begin{figure}
  \includegraphics[width=0.35\textwidth]{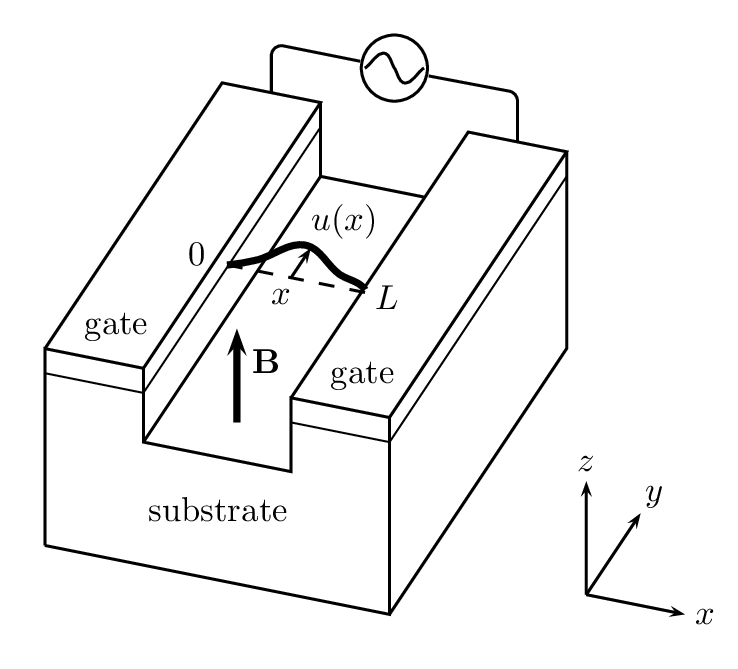}
  \caption{Schematic figure of the setup. A nanowire (thick solid line) is suspended between two metallic gates and oscillates about its equilibrium position. It is influenced by a perpendicular magnetic field $\mathbf{B}$.}
  \label{fig:setup}
\end{figure}

\begin{figure}
  \includegraphics[width=\textwidth]{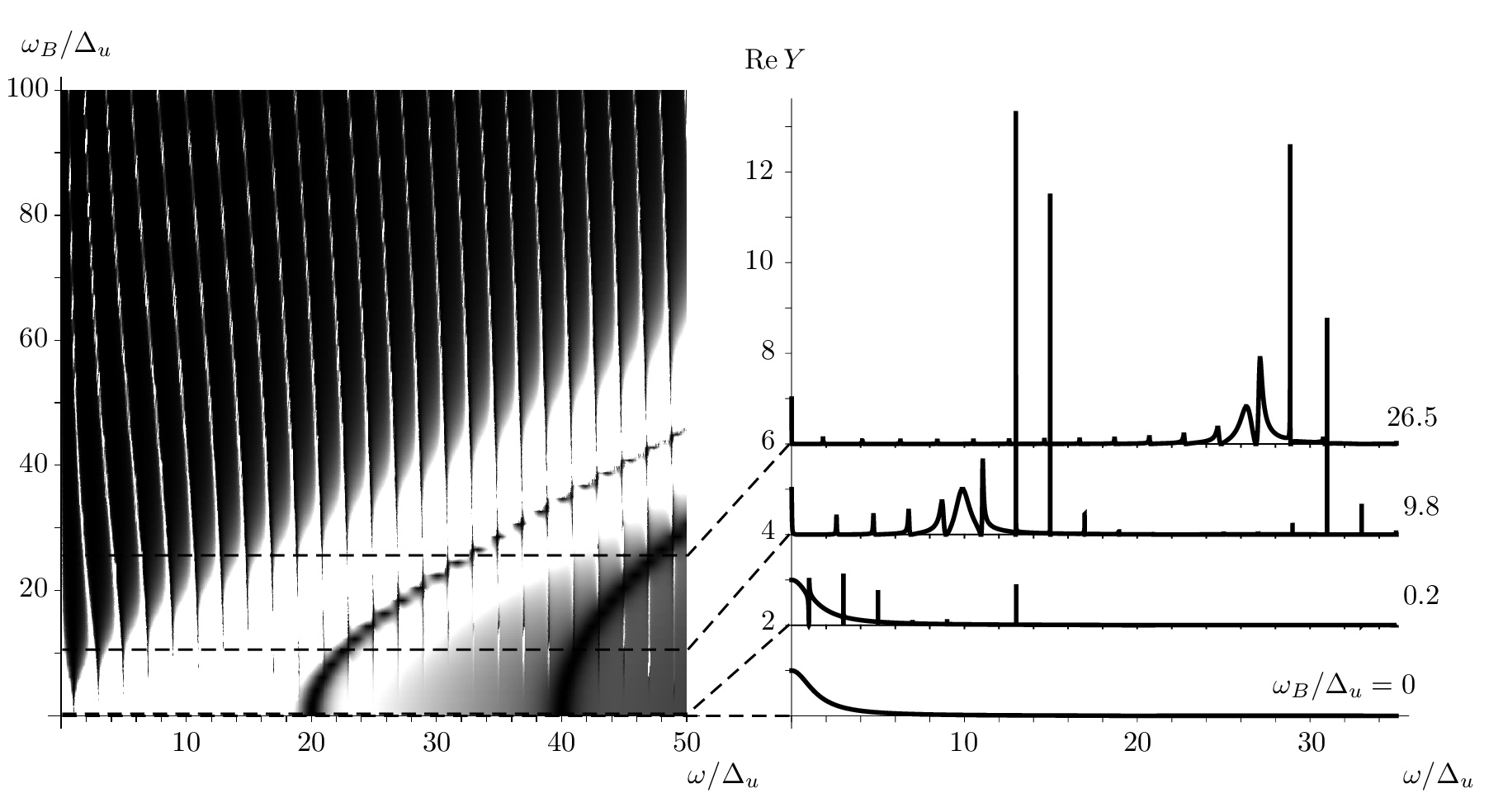}
  \caption{Left panel shows the density plot of the real part of the admittance $\mathrm{Re}\,Y$ as a function of $\omega/\Delta_u$ and $\omega_B/\Delta_u$ where $\omega_B=B\sqrt{v_Fe^2/\pi\hbar\rho c^2}$, and $\Delta_u=\pi\hbar v_s/L$, and $\omega$ is the ac bias frequency. The parameters used here are $\eta=\Delta_u/\Delta_\theta=v_s/v=0.1$ and $K=0.22$. Lighter regions correspond to higher values. The curves in the right panel show cross sections of the density plot at several different magnetic fields; from bottom to top, $\omega_B/\Delta_u = 0,\ 0.2,\ 9.8$, and $26.5$. Curves are offset for better viewing.}
  \label{fig:Y(w)}
\end{figure}
\end{center}


\begin{thebibliography}{19}
\expandafter\ifx\csname natexlab\endcsname\relax\def\natexlab#1{#1}\fi
\expandafter\ifx\csname bibnamefont\endcsname\relax
  \def\bibnamefont#1{#1}\fi
\expandafter\ifx\csname bibfnamefont\endcsname\relax
  \def\bibfnamefont#1{#1}\fi
\expandafter\ifx\csname citenamefont\endcsname\relax
  \def\citenamefont#1{#1}\fi
\expandafter\ifx\csname url\endcsname\relax
  \def\url#1{\texttt{#1}}\fi
\expandafter\ifx\csname urlprefix\endcsname\relax\def\urlprefix{URL }\fi
\providecommand{\bibinfo}[2]{#2}
\providecommand{\eprint}[2][]{\url{#2}}

\bibitem[{\citenamefont{Bishop et~al.}(2001)\citenamefont{Bishop, Giles, and
  Gammel}}]{bishop2001a}
\bibinfo{author}{\bibfnamefont{D.}~\bibnamefont{Bishop}},
  \bibinfo{author}{\bibfnamefont{C.~R.} \bibnamefont{Giles}}, \bibnamefont{and}
  \bibinfo{author}{\bibfnamefont{P.}~\bibnamefont{Gammel}},
  \bibinfo{journal}{Phys.\ Today} \textbf{\bibinfo{volume}{54}},
  \bibinfo{pages}{38} (\bibinfo{year}{2001}).

\bibitem[{\citenamefont{Blencowe}(2004)}]{blencowe2004a}
\bibinfo{author}{\bibfnamefont{M.}~\bibnamefont{Blencowe}},
  \bibinfo{journal}{Phys.\ Rep.} \textbf{\bibinfo{volume}{395}}, \bibinfo{pages}{159}
  (\bibinfo{year}{2004}).

\bibitem[{\citenamefont{Ahn and Yi}(2004)}]{ahn2004a}
\bibinfo{author}{\bibfnamefont{K.-H.} \bibnamefont{Ahn}} \bibnamefont{and}
  \bibinfo{author}{\bibfnamefont{H.}~\bibnamefont{Yi}},
  \bibinfo{journal}{Europhys.\ Lett.} \textbf{\bibinfo{volume}{67}},
  \bibinfo{pages}{641} (\bibinfo{year}{2004}).

\bibitem[{\citenamefont{Knobel and Cleland}(2003)}]{knobel2003a}
\bibinfo{author}{\bibfnamefont{R.~G.} \bibnamefont{Knobel}} \bibnamefont{and}
  \bibinfo{author}{\bibfnamefont{A.~N.} \bibnamefont{Cleland}},
  \bibinfo{journal}{Nature} \textbf{\bibinfo{volume}{424}},
  \bibinfo{pages}{291} (\bibinfo{year}{2003}).

\bibitem[{\citenamefont{Schwab and Roukes}(2005)}]{schwab2005a}
\bibinfo{author}{\bibfnamefont{K.~C.} \bibnamefont{Schwab}} \bibnamefont{and}
  \bibinfo{author}{\bibfnamefont{M.~L.} \bibnamefont{Roukes}},
  \textbf{\bibinfo{volume}{58}}, \bibinfo{pages}{36} (\bibinfo{year}{2005}).

\bibitem[{\citenamefont{LaHaye et~al.}(2004)\citenamefont{LaHaye, Buu,
  Camarota, and Schwab}}]{lahaye2004a}
\bibinfo{author}{\bibfnamefont{M.~D.} \bibnamefont{LaHaye}},
  \bibinfo{author}{\bibfnamefont{O.}~\bibnamefont{Buu}},
  \bibinfo{author}{\bibfnamefont{B.}~\bibnamefont{Camarota}}, \bibnamefont{and}
  \bibinfo{author}{\bibfnamefont{K.~C.} \bibnamefont{Schwab}},
  \bibinfo{journal}{Science} \textbf{\bibinfo{volume}{304}},
  \bibinfo{pages}{74} (\bibinfo{year}{2004}).

\bibitem[{\citenamefont{LeRoy et~al.}(2004)\citenamefont{LeRoy, Lemay, Kong,
  and Dekker}}]{leroy2004a}
\bibinfo{author}{\bibfnamefont{B.~J.} \bibnamefont{LeRoy}},
  \bibinfo{author}{\bibfnamefont{S.~G.} \bibnamefont{Lemay}},
  \bibinfo{author}{\bibfnamefont{J.}~\bibnamefont{Kong}}, \bibnamefont{and}
  \bibinfo{author}{\bibfnamefont{C.}~\bibnamefont{Dekker}},
  \bibinfo{journal}{Nature} \textbf{\bibinfo{volume}{432}},
  \bibinfo{pages}{371} (\bibinfo{year}{2004}).

\bibitem[{\citenamefont{Sapmaz et~al.}(2006)\citenamefont{Sapmaz,
  Jarillo-Herroro, Blanter, Dekker, and van~der Zant}}]{sapmaz2006a}
\bibinfo{author}{\bibfnamefont{S.}~\bibnamefont{Sapmaz}},
  \bibinfo{author}{\bibfnamefont{P.}~\bibnamefont{Jarillo-Herroro}},
  \bibinfo{author}{\bibfnamefont{Y.~M.} \bibnamefont{Blanter}},
  \bibinfo{author}{\bibfnamefont{C.}~\bibnamefont{Dekker}}, \bibnamefont{and}
  \bibinfo{author}{\bibfnamefont{H.~S.~J.} \bibnamefont{van~der Zant}},
  \bibinfo{journal}{Phys.\ Rev.\ Lett.} \textbf{\bibinfo{volume}{96}},
  \bibinfo{pages}{026801} (\bibinfo{year}{2006}).

\bibitem[{\citenamefont{Shekhter et~al.}(2006)\citenamefont{Shekhter, Gorelik,
  Glazman, and Johnson}}]{shekhter2006a}
\bibinfo{author}{\bibfnamefont{R.~I.} \bibnamefont{Shekhter}},
  \bibinfo{author}{\bibfnamefont{L.~Y.} \bibnamefont{Gorelik}},
  \bibinfo{author}{\bibfnamefont{L.~I.} \bibnamefont{Glazman}},
  \bibnamefont{and} \bibinfo{author}{\bibfnamefont{M.}~\bibnamefont{Johnson}},
  \bibinfo{journal}{Phys.\ Rev.\ Lett.} \textbf{\bibinfo{volume}{97}},
  \bibinfo{pages}{156801} (\bibinfo{year}{2006}).

\bibitem[{\citenamefont{Yao et~al.}(1999)\citenamefont{Yao, Postma, Balents,
  and Dekker}}]{yao1999a}
\bibinfo{author}{\bibfnamefont{Z.}~\bibnamefont{Yao}},
  \bibinfo{author}{\bibfnamefont{H.~W.~C.} \bibnamefont{Postma}},
  \bibinfo{author}{\bibfnamefont{L.}~\bibnamefont{Balents}}, \bibnamefont{and}
  \bibinfo{author}{\bibfnamefont{C.}~\bibnamefont{Dekker}},
  \bibinfo{journal}{Nature} \textbf{\bibinfo{volume}{402}},
  \bibinfo{pages}{273} (\bibinfo{year}{1999}).

\bibitem[{\citenamefont{Bardeen et~al.}(1957)\citenamefont{Bardeen, Cooper, and
  Schrieffer}}]{bardeen1957a}
\bibinfo{author}{\bibfnamefont{J.}~\bibnamefont{Bardeen}},
  \bibinfo{author}{\bibfnamefont{L.~N.} \bibnamefont{Cooper}},
  \bibnamefont{and} \bibinfo{author}{\bibfnamefont{J.~R.}
  \bibnamefont{Schrieffer}}, \bibinfo{journal}{Phys.\ Rev.}
  \textbf{\bibinfo{volume}{108}}, \bibinfo{pages}{1175} (\bibinfo{year}{1957}).

\bibitem[{\citenamefont{Su et~al.}(1979)\citenamefont{Su, Schrieffer, and
  Heeger}}]{su1979a}
\bibinfo{author}{\bibfnamefont{W.~P.} \bibnamefont{Su}},
  \bibinfo{author}{\bibfnamefont{J.~R.} \bibnamefont{Schrieffer}},
  \bibnamefont{and} \bibinfo{author}{\bibfnamefont{A.~J.}
  \bibnamefont{Heeger}}, \bibinfo{journal}{Phys.\ Rev.\ Lett.}
  \textbf{\bibinfo{volume}{42}}, \bibinfo{pages}{1698} (\bibinfo{year}{1979}).

\bibitem[{\citenamefont{Tomonaga}(1950)}]{tomonaga1950a}
\bibinfo{author}{\bibfnamefont{S.}~\bibnamefont{Tomonaga}},
  \bibinfo{journal}{Prog.\ Theor.\ Phys.} \textbf{\bibinfo{volume}{5}},
  \bibinfo{pages}{544} (\bibinfo{year}{1950}).

\bibitem[{\citenamefont{Luttinger}(1963)}]{luttinger1963a}
\bibinfo{author}{\bibfnamefont{J.~M.} \bibnamefont{Luttinger}},
  \bibinfo{journal}{J. Math. Phys.} \textbf{\bibinfo{volume}{4}},
  \bibinfo{pages}{1154} (\bibinfo{year}{1963}).

\bibitem[{\citenamefont{Haldane}(1981)}]{haldane1981b}
\bibinfo{author}{\bibfnamefont{F.~D.~M.} \bibnamefont{Haldane}},
  \bibinfo{journal}{Phys.\ Rev.\ Lett.} \textbf{\bibinfo{volume}{47}},
  \bibinfo{pages}{1840} (\bibinfo{year}{1981}).

\bibitem[{\citenamefont{Kane and Fisher}(1992{\natexlab{a}})}]{kane1992a}
\bibinfo{author}{\bibfnamefont{C.~L.} \bibnamefont{Kane}} \bibnamefont{and}
  \bibinfo{author}{\bibfnamefont{M.~P.~A.} \bibnamefont{Fisher}},
  \bibinfo{journal}{Phys.\ Rev.\ Lett.} \textbf{\bibinfo{volume}{68}},
  \bibinfo{pages}{1220} (\bibinfo{year}{1992}{\natexlab{a}}).

\bibitem[{\citenamefont{Yi}(2002)}]{yi2002a}
\bibinfo{author}{\bibfnamefont{H.}~\bibnamefont{Yi}}, \bibinfo{journal}{Phys.\
  Rev.\ B} \textbf{\bibinfo{volume}{65}}, \bibinfo{pages}{195101}
  (\bibinfo{year}{2002}).

\bibitem[{\citenamefont{Maslov and Stone}(1995)}]{maslov1995a}
\bibinfo{author}{\bibfnamefont{D.~L.} \bibnamefont{Maslov}} \bibnamefont{and}
  \bibinfo{author}{\bibfnamefont{M.}~\bibnamefont{Stone}},
  \bibinfo{journal}{Phys.\ Rev.\ B} \textbf{\bibinfo{volume}{52}},
  \bibinfo{pages}{5539} (\bibinfo{year}{1995}).

\bibitem[{\citenamefont{Kane and Fisher}(1992{\natexlab{b}})}]{kane1992b}
\bibinfo{author}{\bibfnamefont{C.~L.} \bibnamefont{Kane}} \bibnamefont{and}
  \bibinfo{author}{\bibfnamefont{M.~P.~A.} \bibnamefont{Fisher}},
  \bibinfo{journal}{Phys.\ Rev.\ B} \textbf{\bibinfo{volume}{46}},
  \bibinfo{pages}{15233} (\bibinfo{year}{1992}{\natexlab{b}}).

\end{thebibliography}
\end{document}